\documentclass[
aps,
prl,
superscriptaddress,
nofootinbib,
twocolumn,
floatfix]{revtex4-1}
\usepackage{amsmath}
\usepackage{amssymb}
\usepackage{graphicx}
\usepackage{dcolumn}
\usepackage{bm}
\usepackage[mathlines]{lineno}

\usepackage{times}
\usepackage{hyperref}

\newcommand{\liga}{LiGaCr$_4$S$_8$}

\begin{document}

\preprint{APS/123-QED}


\title{Cluster Frustration in the Breathing Pyrochlore Magnet \liga{}}

\author{Ganesh Pokharel}
\email{gpokhare@vols.utk.edu}
\affiliation{Department of Physics \& Astronomy, University of Tennessee, Knoxville, TN 37996, USA}
\affiliation{Materials Science \& Technology Division, Oak Ridge National Laboratory, Oak Ridge, TN 37831, USA}

\author{Hasitha Suriya Arachchige}
\affiliation{Department of Physics \& Astronomy, University of Tennessee, Knoxville, TN 37996, USA}
\affiliation{Materials Science \& Technology Division, Oak Ridge National Laboratory, Oak Ridge, TN 37831, USA}

\author{Travis J. Williams}
\affiliation{Neutron Scattering Division, Oak Ridge National Laboratory, Oak Ridge, TN 37831, USA}

\author{Andrew F. May}
\affiliation{Materials Science \& Technology Division, Oak Ridge National Laboratory, Oak Ridge, TN 37831, USA}

\author{Randy S. Fishman}
\affiliation{Materials Science \& Technology Division, Oak Ridge National Laboratory, Oak Ridge, TN 37831, USA}

\author{Gabriele Sala}
\affiliation{Neutron Scattering Division, Oak Ridge National Laboratory, Oak Ridge, TN 37831, USA}

\author{Stuart Calder}
\affiliation{Neutron Scattering Division, Oak Ridge National Laboratory, Oak Ridge, TN 37831, USA}

\author{Georg Ehlers}
\affiliation{Neutron Technologies Division, Oak Ridge National Laboratory, Oak Ridge, TN 37831, USA} 

\author{David S. Parker}
\affiliation{Materials Science \& Technology Division, Oak Ridge National Laboratory, Oak Ridge, TN 37831, USA}

\author{Tao Hong}
\affiliation{Neutron Scattering Division, Oak Ridge National Laboratory, Oak Ridge, TN 37831, USA}

\author{Andrew Wildes}
\affiliation{Institut Laue-Langevin, CS 20156, 38042 Grenoble Cédex 9, France}

\author{David Mandrus}
\affiliation{Department of Materials Science \& Engineering, University of Tennessee, Knoxville, TN 37996, USA}
\affiliation{Materials Science \& Technology Division, Oak Ridge National Laboratory, Oak Ridge, TN 37831, USA}
\affiliation{Department of Physics \& Astronomy, University of Tennessee, Knoxville, TN 37996, USA}

\author{Joseph A. M. Paddison}
\email{paddisonja@ornl.gov}
\affiliation{Materials Science \& Technology Division, Oak Ridge National Laboratory, Oak Ridge, TN 37831, USA}

\author{Andrew D. Christianson}
\email{christiansad@ornl.gov}
\affiliation{Materials Science \& Technology Division, Oak Ridge National Laboratory, Oak Ridge, TN 37831, USA}


\date{\today}

\begin{abstract}
We present a comprehensive neutron scattering study of the breathing pyrochlore magnet \liga. We observe an unconventional magnetic excitation spectrum with a separation of high and low-energy spin dynamics in the correlated paramagnetic regime above a spin-freezing transition at 12(2)\,K. By fitting to magnetic diffuse-scattering data, we  parameterize the spin Hamiltonian. We find that interactions are ferromagnetic within the large and small tetrahedra of the breathing pyrochlore lattice, but antiferromagnetic further-neighbor interactions are also essential to explain our data, in qualitative agreement with density-functional theory predictions [Ghosh \emph{et al.}, \emph{npj Quantum Mater.} \textbf{4}, 63 (2019)]. We explain the origin of geometrical frustration in \liga{} in terms of net antiferromagnetic coupling between emergent tetrahedral spin clusters that occupy a face-centered lattice. Our results provide insight into the emergence of frustration in the presence of strong further-neighbor couplings, and a blueprint for the determination of magnetic interactions in classical spin liquids.

\end{abstract}

\pacs{Valid PACS appear here}
\maketitle

Geometrical frustration---the inability to satisfy all interactions simultaneously due to geometrical constraints---can generate unusual magnetic states in which long-range magnetic ordering is suppressed but strong short-range spin correlations endure \cite{Balents_2010}. Canonical models of frustrated magnetism often consider spins coupled by antiferromagnetic nearest-neighbor (NN) interactions, which generate a macroscopic degeneracy of magnetic ground states on lattices such as the pyrocholore network of corner-sharing tetrahedra \cite{Moessner_1998,Moessner_1998a,Canals_1998}. This ground-state degeneracy is not symmetry-protected, and in general is expected to be broken by perturbations such as further-neighbor interactions or spin-lattice coupling. Remarkably, however, some materials exhibit highly-frustrated behavior, despite having complex magnetic interactions that deviate strongly from canonical frustrated models \cite{Balz_2016,Paddison_2013a,Venderbos_2011}. These states are of fundamental interest because they can reveal novel frustration mechanisms.

\begin{figure}
\centering
\includegraphics[width=8.5cm]{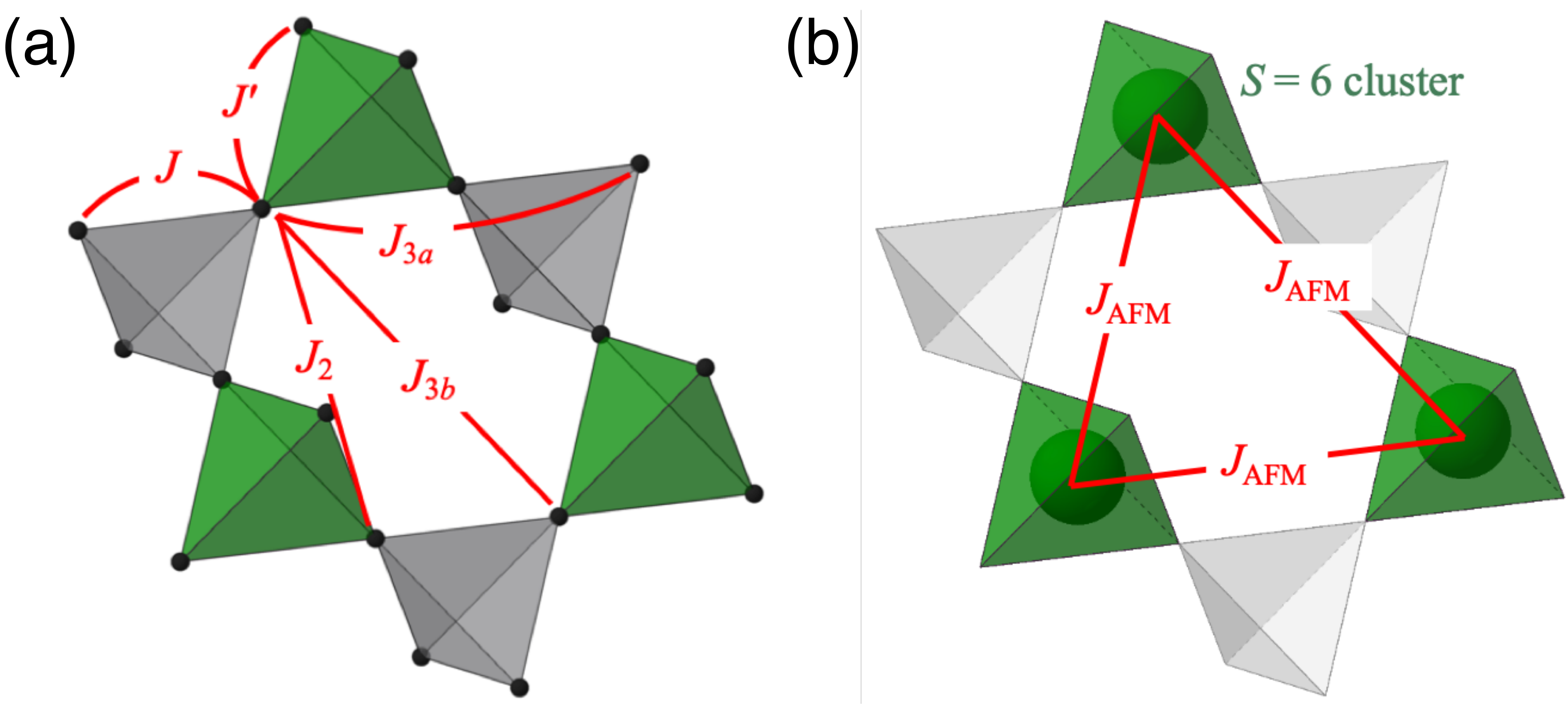}

    \caption{ 
    (a) Breathing pyrochlore lattice of $S=3/2$ Cr$^{3+}$ ions (black circles) in \liga{}, showing large (small) tetrahedra (colored green (grey)), and the connectivity of the exchange interactions $J$, $J^{\prime}$, $J_2$, $J_{3a}$, and $J_{3b}$. $J_{3a}$ and $J_{3b}$ span the same distance but have different symmetry. (b) Emergent tetrahedral clusters generated by strong ferromagnetic $J^{\prime}$ interactions, coupled by a net antiferromagnetic interaction $J_{\mathrm{AFM}}\propto J+4J_{2}+2J_{3a}+2J_{3b}$ ($>0$).    }
\label{fig1}
\end{figure}

A modification of the pyrochlore lattice with potential to realize such states is an alternating array of small and large tetrahedra [Fig.~\ref{fig1}(a)]. This lattice is conventionally called a ``breathing pyrochlore", although the size alternation is static, and corresponds to a symmetry lowering from $Fd\bar{3}m$ to $F\bar{4}3m$ \cite{Okamoto_2013}. Different exchange interactions can occur within the small and large tetrahedra ($J$ and $J^\prime$, respectively; see Fig.~\ref{fig1}(a)), increasing the richness of the phase diagram \cite{Benton_2015}. Neglecting further-neighbor interactions,  conventional ordering is expected only if both $J$ and $J^\prime$ are ferromagnetic. If $J$ and $J^\prime$ are both antiferromagnetic, the ground state is a classical spin liquid, whereas if $J$ and $J^\prime$ are of opposite sign, the ground-state manifold is dimensionally reduced \cite{Benton_2015}. Further-neighbor interactions ($J_2$, $J_{3a}$, and $J_{3b}$; see Fig.~\ref{fig1}(a)) can generate further exotic phases. Perhaps the most intriguing of these is predicted \cite{Ghosh_2019} to occur when $J$ or $J^\prime$ is large and ferromagnetic, and further-neighbor interactions are antiferromagnetic. The dominant ferromagnetic interactions drive the formation of ferromagnetic tetrahedral clusters, and inter-cluster interactions are frustrated because these clusters occupy a face-centered cubic (FCC) lattice [Fig.~\ref{fig1}(b)] \cite{Ghosh_2019}. This model provides a notable example of the concept of emergent frustration---the frustration of multi-spin degrees of freedom that occupy a different lattice to the spins themselves \cite{Paddison_2013a,Venderbos_2011}.

\begin{figure*}[htp]
\centering
\includegraphics[]{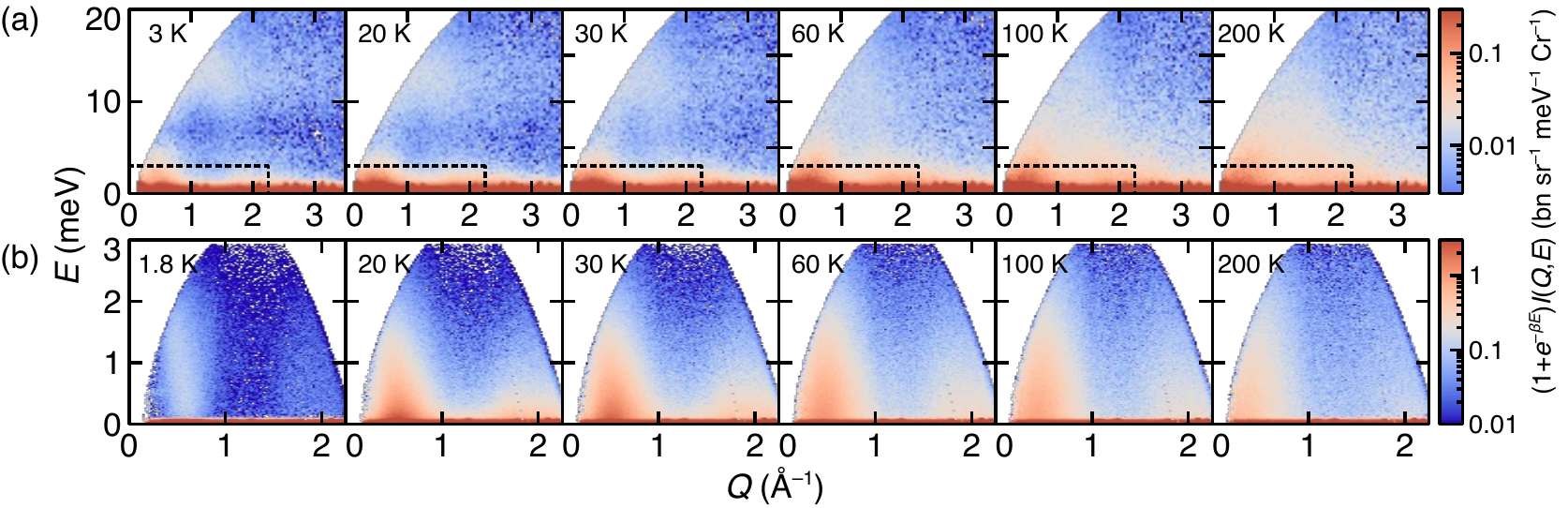}
    \caption{Inelastic neutron scattering spectra of \liga {} measured at temperatures indicated in the panels. (a) High energy excitation spectra measured with $E_i=25$\,meV. (b) Low energy spin excitations measured with $E_i = 3.32$\,meV. The regions of $Q$-$E$ space enclosed by dotted lines in (a) indicate the regions shown in (b). Intensity is corrected for detailed balance and shown by color on a logarithmic scale. The intensity scale in (b) is a factor of 10 larger than in (a).
    }

\label{fig2}
\end{figure*}

Experimental realizations of the breathing pyrochlore model include the spinel derivatives $AA^{\prime}$Cr$_{4}X_{8}$, in which the $A$-site is occupied by an ordered arrangement of Li$^+$ and In$^{3+}$/Ga$^{3+}$; $X=\textrm{O, S or Se}$; and the Cr$^{3+}$ ions occupy a breathing pyrochlore lattice \cite{Okamoto_2013}. Since $J\sim J^{\prime}$ in these materials, collective magnetic behavior is expected, in contrast to the breathing pyrochlore material Ba$_3$Yb$_2$Zn$_5$O$_{11}$ in which tetrahedra are decoupled \cite{Kimura_2014,Haku_2016,Rau_2018,Rau_2016,Park2016}. Series members with $X=\textrm{O}$ have antiferromagnetic  $J$ and $J^{\prime}$ and exhibit magnetostructural phase transitions and nematic spin ordering \cite{Okamoto_2013,Okamoto_2015,Lee_2016,Wawrzynczak_2017,Tanaka_2018}. Replacement of O with S or Se ligands is predicted to cause two key differences: suppression of direct exchange relative to superexchange, which is expected to be ferromagnetic because the Cr--X--Cr bond angles are near to 90$^{\circ}$ \cite{Pokharel_2018}; and enhancement of further-neighbor interactions \cite{Ghosh_2019}. Hence, series members with S or Se ligands \cite{plumier1971,plumier1977,Pokharel_2018,Okamoto_2018,Duda_2008} are promising candidates to realize models of frustration driven by further-neighbor interactions. However, no experimental determination of the magnetic interactions in such systems exists.

Here, we use neutron scattering measurements to study the breathing pyrochlore LiGaCr$_{4}$S$_{8}$. While the Weiss constant of LiGaCr$_{4}$S$_{8}$ is relatively small, $\theta_{\mathrm{CW}} \approx 20$\,K \cite{PINCH_1970, Okamoto_2018, Pokharel_2018}, its bulk magnetic susceptibility $\chi$ shows strong deviations from Curie-Weiss behavior below $\sim$100\,K, suggesting the development of strong spin correlations above its spin-freezing transition at $T_f = 12(2)$\,K \cite{Pokharel_2018}. Spin freezing is probably driven by a small amount of off-stoichiometry, as approximately 4\% of Li sites are occupied by Ga \cite{Pokharel_2018}. Our three key results explain the nature and origin of spin correlations in LiGaCr$_{4}$S$_{8}$: we experimentally parameterize the spin Hamiltonian to reveal the importance of further-neighbor couplings; we confirm recent theoretical predictions (Ref.~\onlinecite{Ghosh_2019}) of cluster frustration; and we observe a direct signature of cluster formation in its magnetic excitation spectrum. These results show that LiGaCr$_{4}$S$_{8}$ realizes the frustration of tetrahedral clusters on an emergent FCC lattice.

Fig.~\ref{fig2} presents the temperature dependence of our inelastic neutron scattering (INS) data as a function of wavevector transfer $Q=|\mathbf{Q}|$ and energy transfer $E$. Data were collected on a $\sim$2\,g polycrystalline sample (see supplementary material (SM)\cite{supp}) using two neutron spectrometers: Fig.~\ref{fig2}(a) shows high-energy data measured using the ARCS spectrometer with incident energy $E_i = 25$\,meV, and Fig.~\ref{fig2}(b) shows low-energy data measured using the CNCS spectrometer with  $E_i = 3.32$\,meV. All INS data have been corrected for detailed balance, and CNCS data are background-subtracted.
The dependence of the scattering on $Q$ and temperature suggests that it is of magnetic origin. 

\begin{figure*}
\centering
\includegraphics[]{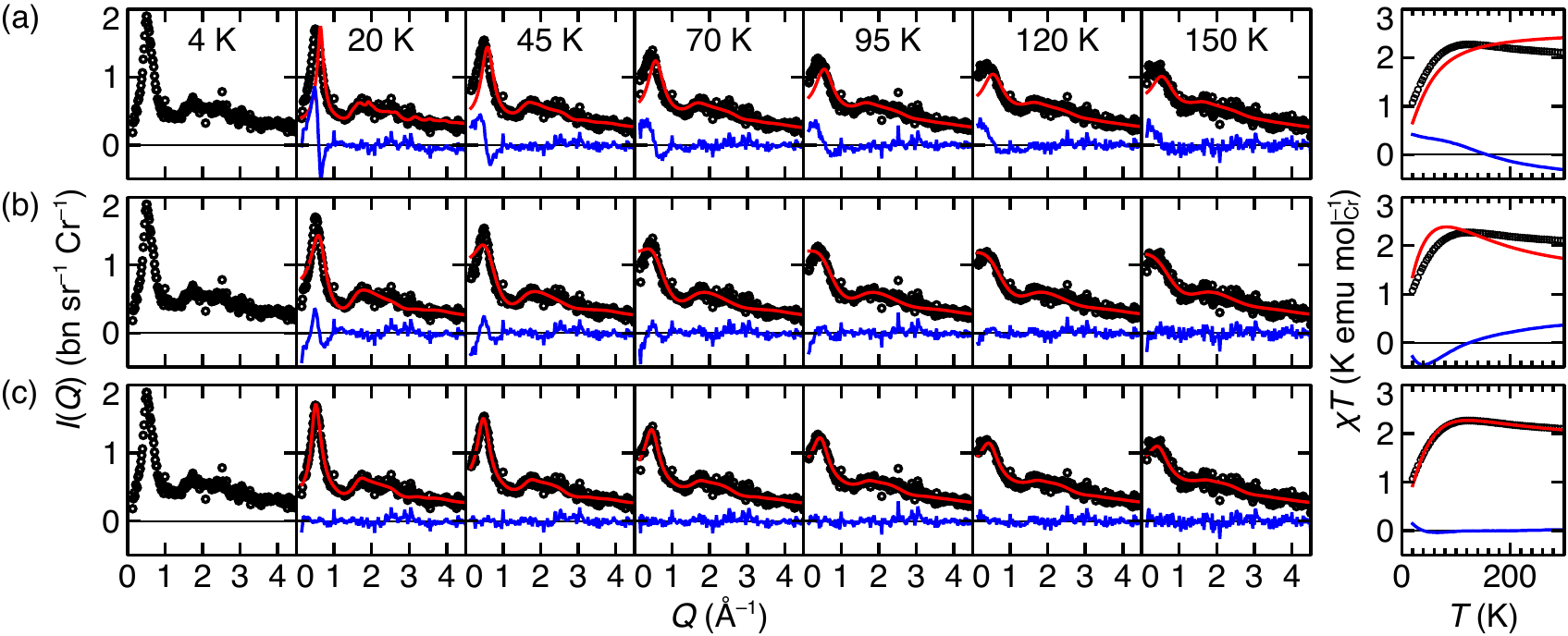}
    \caption[width=2\linewidth]{Data (black circles), model fits (red lines), and data--fit (blue lines) for (a) the DFT model of Ref. [\onlinecite{Ghosh_2019}], (b) the $J$-$J^{\prime}$ model, and (c) the $J$-$J^{\prime}$-$J_2$-$J_{3a}$-$J_{3b}$ model discussed in the text. The left-hand panels of (a), (b) and (c) represent the neutron scattering data at temperatures indicated in each panel, and the right-hand panel represents $\chi$. Fits were performed for $T\geq 20$\,K. 
    }
\label{fig5}
\end{figure*}

The bandwidth of the spectrum is about $15$\,meV, which is larger than $\theta_\mathrm{CW} \approx 20$\,K ($2$\,meV),  
suggesting that both ferromagnetic and antiferromagnetic exchange interactions are significant.
Above $100$\,K the spectrum is broad, as expected for a paramagnet. In contrast, between $20$\,K and $100$\,K, a band at 12 meV and low-energy quasielastic excitations are observed. The low-energy scattering is much more intense than the high-energy scattering, and has a pronounced wavevector dependence. On cooling, the quasielastic scattering moves towards low energy; however, analysis of the dynamical susceptibility using a damped-harmonic-oscillator model indicates that the excitations are overdamped and gapless at all measured temperatures\cite{supp}. 
Below $T_f$, most of the quasielastic spectral weight shifts to the elastic line\cite{supp}, consistent with the expected dramatic slowing-down of spin dynamics associated with spin freezing \cite{young1986}. Interestingly, the intensity of the high-energy band does not change appreciably compared to $20$\,K---a point to which we return below. Additional evidence of spin freezing is provided by our muon spin relaxation ($\mu$SR) measurements, described in detail in the SM \cite{supp}.  Zero-field $\mu$SR measurements down to 1.8~K showed no evidence of static magnetic order or a canonical spin glass state, however the relaxation rate increased at the same temperature as seen with neutron scattering suggesting a slowing down of the spin fluctuations towards a frozen magnetic state.  Longitudinal-field $\mu$SR does not show evidence of dynamic spin fluctuations, but rather agrees with the emergence of spin freezing at low temperature in \liga{}.

\begin{table}
\centering
\caption[width=1\columnwidth]{Magnetic interaction parameters for different models. Parameter values held fixed are denoted by an asterisk (\**). }
\begin{tabular}{c c c c c c c}
\hline\hline
Model & $\it{J}$ (K) & $\it{J^\prime}$ (K) & $\it{J_2}$ (K) & $\it{J_{3a}}$ (K) & $\it{J_{3b}}$ (K) \\ [0.5ex]
\hline\hline
DFT (Ref.~\onlinecite{Ghosh_2019}) & $-7.7(1)$ & $-12.2(1)$ & $1.2(1)$& $6.1(1)$ & $3.0(1)$ \\
$J$-$J^\prime$ & $3.07(3)$ & $-29.9(4)$ & $0$\**& $0$\** & $0$\** \\ 
$J$-$J^\prime$-$J_2$-$J_{3a}$-$J_{3b}$ & $-7.8(6)$ & $-22.1(3)$ & $-1.6(4)$& $9.6(1)$ & $0.8(4)$ \\ [1ex]
\hline
\end{tabular}
\label{table}
\end{table}

We now obtain an estimate of the magnetic interactions in LiGaCr$_4$S$_8$. Our starting point is a Heisenberg spin Hamiltonian, $H=\frac{1}{2}\sum_{i,j}J_{ij}\mathbf{S}_i \cdot \mathbf{S}_j,$
which has been applied successfully to Cr$^{3+}$-based spinels \cite{Bai_2019,Matsuda_2007}. Here, 
 $J_{ij}\in \{J,J^{\prime},J_2,J_{3a},J_{3b}\}$ denotes an interaction as shown in Fig.~\ref{fig1}, and $\mathbf{S}$ denotes a classical vector of magnitude $\sqrt{S(S+1)}$ with $S = 3/2$.  Because LiGaCr$_4$S$_8$ does not exhibit long-range magnetic order, it is not possible to employ the conventional approach of fitting interactions to spin-wave spectra in an ordered state. Therefore, we consider instead the magnetic diffuse scattering intensity, $I(Q)=\int I(Q,E)\thinspace dE$, which we obtain from background-corrected powder-diffraction data collected using the HB-2A diffractometer at ORNL (see SM \cite{supp}). For a given set of interaction parameters, we calculate $I(Q)$ and$\chi T$ using Onsager reaction field theory \cite{Brout_1967,Logan_1995,Hohlwein_2003}, which is equivalent to the self-consistent Gaussian approximation used elsewhere \cite{Benton_2015,Bai_2019,Plumb_2019} and gives accurate results for frustrated Heisenberg pyrochlore models \cite{Conlon_2010}.

We tested three models against our $I(Q)$ data and the $\chi T$ data from Ref.~\onlinecite{Pokharel_2018} [Fig.~\ref{fig5}]. Values of the interaction parameters for each model are given in Table~\ref{table}. First, we considered the five-parameter ``DFT model" obtained using density-functional theory (DFT) in Ref.~\onlinecite{Ghosh_2019}. Calculations of $I(Q)$ and $\chi T$ for this model show partial agreement with experiment; however, the calculated position of the main diffuse-scattering peak disagrees with the data [Fig.~\ref{fig5}(a)]. Second, we fitted a simpler model to $I(Q)$ data that included $J$ and $J^\prime$ interactions only (``$J$-$J^\prime$ model"). These fits also do not agree with the $I(Q)$ data, and are inconsistent with the $\chi T$ data [Fig.~\ref{fig5}(b)]. Crucially, this result indicates that longer-ranged interactions beyond $J$ and $J^\prime$ are essential to account for our experimental data. Finally, we fitted all five interaction parameters to our $I(Q)$ and $\chi T$ data (``$J$-$J^\prime$-$J_2$-$J_{3a}$-$J_{3b}$ model"). Our data robustly determine a unique optimal solution (see SM \cite{supp}) that gives a good fit to $I(Q)$ and $\chi T$ [Fig.~\ref{fig5}(c)]. We find $J^{\prime}$ is the largest interaction, $J$, $J^\prime$ are ferromagnetic, $J_{3a}$ is antiferromagnetic, and $J_2$ and $J_{3b}$ are small. The DFT model \cite{Ghosh_2019} shows the same trends.  The consistency between the results derived by the two methods suggests that the trends determined by the modeling are physically reasonable.

\begin{figure}
\centering
\includegraphics[width=8.5cm]{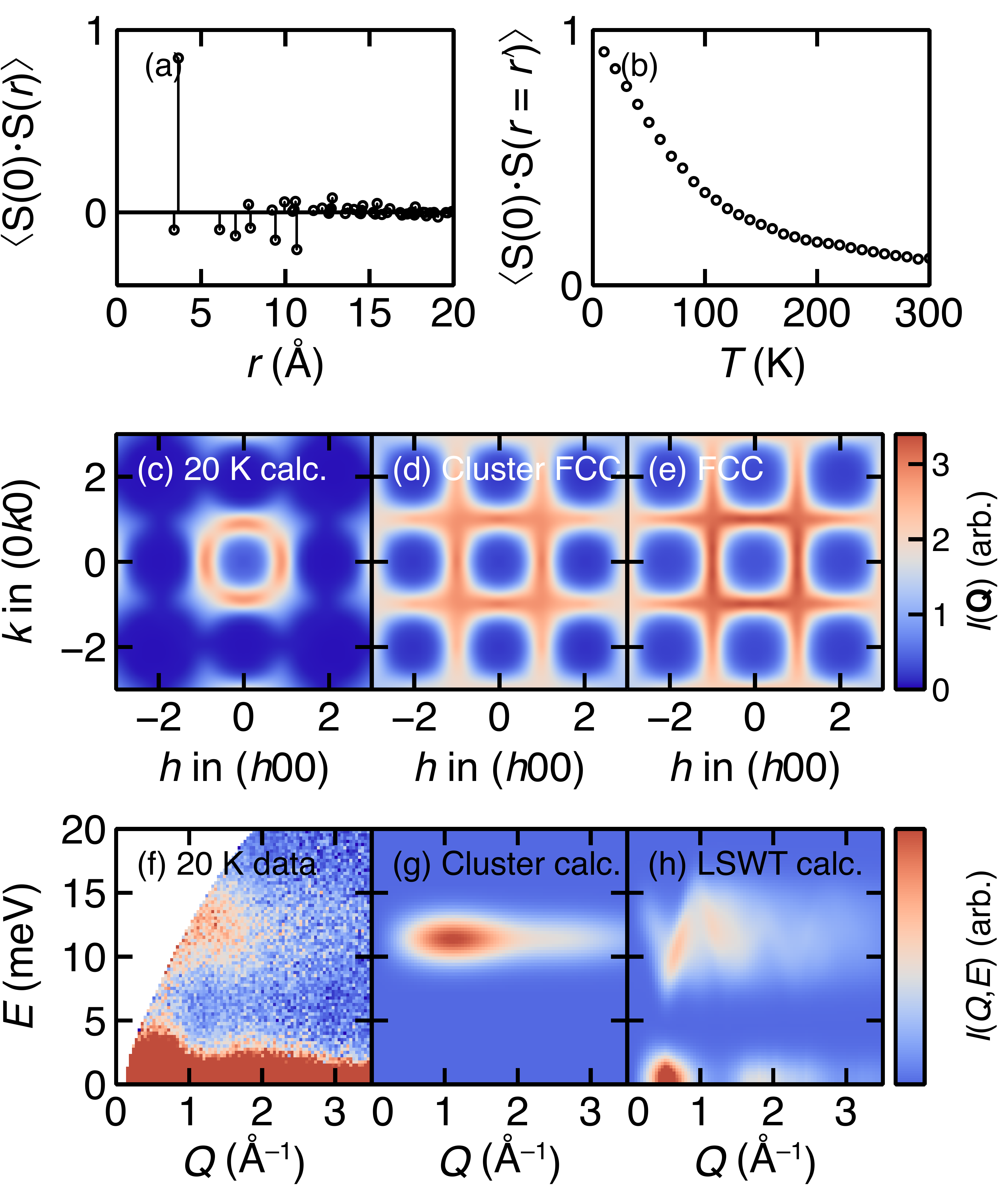}
    \caption{(a) Spin-pair correlation $\langle\mathbf{S}(0)\cdot\mathbf{S}(r)\rangle $ as a function of distance $r$ between spins. The normalization is such that $\langle\mathbf{S}(0)\cdot\mathbf{S}(0)\rangle =1$. Results were obtained from classical Monte Carlo simulations driven by interaction parameters optimized to our neutron data. (b) Calculated temperature dependence of the spin-spin correlation function at $r=r^{\prime}$, the distance between neighboring spins in large tetrahedra. (c) Calculated single-crystal diffuse-scattering pattern $I(\mathbf{Q})$ in the $(hk0)$ plane at 20\,K. (d) Calculated $I(\mathbf{Q})$ from the emergent FCC lattice of cluster spins, defined as $\mathbf{S}^{\prime}=\sum_{i=1}^{4}\mathbf{S}_{i}$ on each tetrahedron. (e) Calculated $I(\mathbf{Q})$ for spins on the FCC lattice with antiferromagnetic NN exchange interactions $J_\mathrm{AFM}= 0.43$\,K. (f) Experimentally-measured spin excitation spectrum at 20\,K. (g) Calculated spin excitation spectrum of an isolated tetrahedral cluster. (h) Spin excitation spectrum calculated assuming a proximate ordered ground state with propagation vector $\mathbf{k}=[0,0,1]$ \cite{Toth_2015}. 
}
\label{fig6}
\end{figure}

With an interaction model in hand, we consider the origin of frustration in LiGaCr$_4$S$_8$. We hypothesize that, at low temperature, spins coupled by dominant ferromagnetic $J^{\prime}$ are essentially aligned within the large tetrahedra, forming $S^{\prime}\approx6$ clusters. The lattice occupied by these clusters is FCC [Fig.~\ref{fig1}(b)], and the net interaction between  clusters for our parameters is given by $J_{\mathrm{AFM}}=(J+4J_{2}+2J_{3a}+2J_{3b})/16=0.43$\,K \cite{Ghosh_2019}; i.e., it is antiferromagnetic. We therefore also hypothesize the suppression of $T_f$ compared to $J^{\prime}$ occurs because of the frustration of antiferromagnetic inter-cluster interactions on the FCC lattice, as proposed theoretically in Ref. [\onlinecite{Ghosh_2019}].

To test the hypothesis of ferromagnetic cluster formation, we performed classical Monte Carlo simulations driven by our fitted interaction parameters (see SM \cite{supp}). Fig.~\ref{fig6}(a) shows that, at $20$\,K, the simulated spin correlation function $\langle\mathbf{S}(0)\cdot\mathbf{S}(r)\rangle $ is close to unity at the distance $r^{\prime}$ within large tetrahedra. This result shows that large tetrahedral clusters are essentially ferromagnetic at $20$\,K. Fig.~\ref{fig6}(b) shows the calculated temperature dependence of $\langle\mathbf{S}(0)\cdot\mathbf{S}(r=r^{\prime})\rangle $, and reveals that the clusters develop below $100$\,K. 
As described in the SM \cite{supp}, our own all-electron first principles calculations  support the “ferromagnetic cluster” picture presented here, along with the counterintuitive distance dependence of the exchange interactions.
To test the hypothesis of antiferromagnetic frustration of $S^{\prime}\approx6$ cluster spins, we calculated the Fourier transform of the 3D spin correlation function $I(\mathbf{Q})\propto \sum_{\mathbf{r}} \left\langle \mathbf{S}(\mathbf{0})\cdot\mathbf{S}(\mathbf{r})\right\rangle \exp(i\mathbf{Q}\cdot\mathbf{r})$ from our Monte Carlo model using the program Scatty \cite{Paddison_2019}. 
Fig.~\ref{fig6}(c) shows the calculated $I(\mathbf{Q})$ for \liga~at $20$\,K. 
Fig.~\ref{fig6}(d) shows the calculated $I(\mathbf{Q})$ from the FCC lattice of “cluster spins”, defined on each tetrahedron of the breathing-pyrochlore lattice as $\mathbf{S}^{\prime}=\sum_{i=1}^{4}\mathbf{S}_{i}$. Figs.~\ref{fig6}(c) and (d) are different because the former includes the structure factor of the tetrahedral cluster, whereas the latter does not.
Fig.~\ref{fig6}(e) shows the calculated $I(\mathbf{Q})$ for spins on the FCC lattice coupled by NN interactions $J_{\mathrm{AFM}}$. The strong similarity between Figs.~\ref{fig6}(d) and (e) demonstrates that antiferromagnetic interactions between cluster spins in LiGaCr$_4$S$_8$ are frustrated in the same way as individual spins on the FCC lattice. 

The cluster model helps to explain our INS data. Our $20$\,K data are shown on a linear  scale in Fig.~\ref{fig6}(f). 
From ferromagnetic-cluster spin-wave theory \cite{Prsa_2018,Toth_2015}, we calculate that the excitation spectrum of an isolated tetrahedron with interaction $J^{\prime}$ contains a single flat mode at $E=4J^{\prime}S$, whose intensity shows a broad peak centered at a $Q$ of approximately $1.1$\,\AA$^{-1}$ [Fig.~\ref{fig6}(g)]. Despite the simplicity of this calculation, it is in qualitative agreement with both the energy and wavevector dependence of the high-energy excitation in our INS data. 
The single-cluster approximation neglects the effect of coupling between the tetrahedra and consequently contains no low-energy excitations. A different approximation is obtained by optimizing an ordered magnetic ground state using the SpinW software \cite{Toth_2015}: this state again has ferromagnetic spins within large tetrahedra, but are ordered with propagation vector $\mathbf{k}=[0,0,1]$. The assumption of an ordered ground state proximate to the $20$\,K state allows the spectrum to be calculated from linear spin-wave theory, but overestimates the effect of coupling between tetrahedra [Fig.~\ref{fig6}(h)].

Our determination of the magnetic interactions of the breathing-pyrochlore magnet LiGaCr$_4$S$_8$ sets a benchmark for quantitative interpretation of neutron data from polycrystalline samples. Our results show that further-neighbor interactions are large, in agreement with DFT predictions \cite{Ghosh_2019} but in sharp contrast to oxide spinels \cite{Bai_2019}.
The origin of frustration in LiGaCr$_4$S$_8$  is the formation of tetrahedral clusters due to a dominant ferromagnetic $J^{\prime}$ interaction, and the frustration of net antiferromagnetic inter-cluster interactions. We directly observe cluster formation \emph{via} the development of an essentially intra-cluster high-energy mode in INS data. Such modes may potentially be present in other materials where emergent clusters are coupled by frustrated interactions, such as the quantum-spin-liquid candidate Ca$_{10}$Cr$_7$O$_{28}$ \cite{Balz_2016} and the metallic frustrated magnet $\beta$-Mn$_{0.8}$Co$_{0.2}$ \cite{Paddison_2013a}.
Intriguingly, on traversing $T_f$, the high-energy mode in our INS data remains unchanged, whereas the low-energy excitations shift to the elastic line. Hence, the timescale of inter-cluster dynamics is enhanced below $T_f$, while that of the intra-cluster dynamics is unchanged. 
From the frequency dependence of ac susceptibility data \cite{Pokharel_2018}, we obtained the Mydosh parameter $\delta T_f \sim 0.012$ (see SM \cite{supp}). This value
is an order of magnitude larger than that of canonical spin-glass systems such as AuMn \cite{Mulder_1982} and CuMn \cite{Mulder_1981}, but is compatible with  cluster-glass systems such as Cr$_{0.5}$Fe$_{0.5}$Ga \cite{Bag_2018} and Zn$_3$V$_3$O$_8$ \cite{Chakrabarty_2014}, suggesting that the ground state of LiGaCr$_4$S$_8$ is cluster-glass-like.
It would therefore be interesting to investigate whether traditional cluster-glass materials---in which strong structural disorder typically generates clusters with a broad size distribution---exhibit distinct high-energy excitations similar to LiGaCr$_4$S$_8$.


\begin{acknowledgments}
We thank C. Batista for useful discussions and Gerald Morris for technical support with muon spin resonance measurements.  This work was supported by the U.S. Department of Energy, Office of Science, Basic Energy Sciences, Materials Sciences and Engineering Division. G.P. and H.S.A were partially supported by the Gordon and Betty Moore Foundation's EPiQS Initiative through Grant GBMF4416. JAMP acknowledges financial support from Churchill College, University of Cambridge, during early stages of this project. This research used resources at the Spallation Neutron Source and the High Flux Isotope Reactor, a Department of Energy (DOE) Office of Science User Facility operated by Oak Ridge National Laboratory (ORNL).
\end{acknowledgments}

%

\end{document}